\documentclass[lettersize,journal]{IEEEtran}
\usepackage{amsmath,amsfonts}
\usepackage{algorithmic}
\usepackage{algorithm}
\usepackage{array}
\usepackage[caption=false,font=normalsize,labelfont=sf,textfont=sf]{subfig}
\usepackage{textcomp}
\usepackage{stfloats}
\usepackage{url}
\usepackage{verbatim}
\usepackage{graphicx}
\usepackage{cite}
\usepackage{xcolor}
\usepackage{multirow,makecell}

\begin{document}

\title{Secure Digital Semantic Communications: \\ Fundamentals, Challenges, and Opportunities}

\author{
Weixuan Chen,~\IEEEmembership{Graduate Student Member, IEEE}, Qianqian Yang,~\IEEEmembership{Member, IEEE}, Yuanyuan Jia, \\ Junyu Pan,~\IEEEmembership{Graduate Student Member, IEEE}, Shuo Shao, Jincheng Dai,~\IEEEmembership{Member, IEEE}, \\ Meixia Tao,~\IEEEmembership{Fellow, IEEE}, and Ping Zhang,~\IEEEmembership{Fellow, IEEE}


\thanks{

Weixuan Chen, Qianqian Yang$^{\dag}$, Yuanyuan Jia, and Junyu Pan are with the College of Information Science and Electronic Engineering, Zhejiang University, Hangzhou 310027, China. (e-mails: \{weixuanchen, qianqianyang20$^{\dag}$, labulado, junyupan\}@zju.edu.cn).

Shuo Shao is with the Department of System Science, University of Shanghai for Science and Technology, Shanghai 200093, China. (e-mail: shuoshao@usst.edu.cn).

Jincheng Dai is with the Key Laboratory of Universal Wireless Communications, Ministry of Education, Beijing University of Posts and Telecommunications, Beijing 100876, China. (e-mail: daijincheng@bupt.edu.cn).

Meixia Tao is with the Department of Electronic Engineering and the Cooperative Medianet Innovation Center (CMIC), Shanghai Jiao Tong University, Shanghai 200240, China. (e-mail: mxtao@sjtu.edu.cn).

Ping Zhang is with the State Key Laboratory of Networking and Switching Technology, Beijing University of Posts and Telecommunications, Beijing 100876, China. (e-mail: pzhang@bupt.edu.cn).

This work is partly supported by the NSFC under grant No. 62293481, No. 62571487, No. 62201505, by the National Key R\&D Program of China under Grant 2024YFE0200802, and by the Zhejiang Provincial Natural Science Foundation of China under Grant No. LZ25F010001.  (Corresponding author: Qianqian Yang.)

}
}

\maketitle

\begin{abstract}

Semantic communication (SemCom) has emerged as a promising paradigm for future wireless networks by prioritizing task-relevant meaning over raw data delivery, thereby reducing communication overhead and improving efficiency. 
However, shifting from bit-accurate transmission to task-oriented delivery introduces new security and privacy risks. These include semantic leakage, semantic manipulation, knowledge base vulnerabilities, model-related attacks, and threats to authenticity and availability.
Most existing secure SemCom studies focus on analog SemCom, where semantic features are mapped to continuous channel inputs. In contrast, digital SemCom transmits semantic information through discrete bits or symbols within practical transceiver pipelines, offering stronger compatibility with real-world systems while exposing a distinct and underexplored attack surface. 
In particular, digital SemCom typically represents semantic information over a finite alphabet through explicit digital modulation, following two main routes: probabilistic modulation and deterministic modulation. 
These discrete mechanisms and practical transmission procedures introduce additional vulnerabilities affecting bit- or symbol-level semantic information, the modulation stage, and packet-based delivery and protocol operations.
Motivated by these challenges and the lack of a systematic analysis of secure digital SemCom, this paper provides a structured review of the area.
Specifically, we review SemCom fundamentals and clarify the architectural differences between analog and digital SemCom.
We then summarize threats shared by both paradigms and organize the threat landscape specific to digital SemCom, followed by a discussion of potential defenses.
Finally, we outline open research directions toward secure and deployable digital SemCom systems.

\end{abstract}


\section{Introduction}

Next-generation wireless networks are expected to support an unprecedented mix of immersive media, massive sensing, and edge intelligence.
In this context, semantic communication (SemCom) \cite{zhang2024semanticsurvey} has gained increasing attention as a paradigm that improves communication efficiency while preserving task performance. Instead of treating every bit as equally important, SemCom aims to transmit task-relevant meaning or intent, often by exploiting context, side information, or shared knowledge between the transmitter and the receiver. This paradigm has rapidly attracted interest across both the communications and machine learning communities \cite{chaccour2024less}, as it links task-level objectives to measurable communication and system gains.
These properties make SemCom well suited to practical scenarios such as connected autonomous systems, industrial automation and robotics, remote monitoring and healthcare, and extended reality.


However, the characteristics that make SemCom efficient and task-oriented also raise new concerns for security and privacy. Since SemCom often transmits task-oriented representations, the transmitted content may inadvertently reveal sensitive information. In particular, an eavesdropper may not need to reconstruct the source data to obtain useful information, because semantic information can already enable inference of private attributes.
Beyond passive leakage, SemCom is also vulnerable to semantic manipulation. In such attacks, an adversary perturbs the source or semantic information to distort the recovered meaning and mislead downstream tasks, potentially causing misinformation, unsafe decisions, or service disruption.
In addition, SemCom commonly relies on data-driven models and knowledge bases, expanding the attack surface beyond the physical channel. 
Adversaries may compromise the learning pipeline through data poisoning or backdoor insertion, exploit model-related vulnerabilities such as model extraction and inversion, or manipulate the knowledge base to bias semantic interpretation and induce systematic errors. Addressing these security and privacy challenges is therefore essential for practical deployment.

A number of recent works have summarized security and privacy issues in SemCom \cite{2025mengsurvey,yang2024secure,shen2023secure,won2024resource,guo2024survey,do2025security}. In particular, Meng \textit{et al.} \cite{2025mengsurvey} surveyed secure SemCom, cataloged representative attacks, and organized defenses into learning-based, cryptographic, and physical-layer techniques. Yang \textit{et al.} \cite{yang2024secure} and Shen \textit{et al.} \cite{shen2023secure} reviewed fundamental concepts and core challenges for secure SemCom from complementary perspectives. Won \textit{et al.} \cite{won2024resource} discussed security and privacy together with resource management in SemCom, while Guo \textit{et al.} \cite{guo2024survey} examined security and privacy in multi-agent SemCom networks. Do \textit{et al.} \cite{do2025security} further summarized deployment-facing threats and countermeasures.
%
Most existing secure SemCom overviews do not explicitly distinguish between analog and digital SemCom, and instead present a unified discussion of threats and defenses across both paradigms.
Meanwhile, several works have analyzed digital SemCom architectures and enabling techniques in depth \cite{zou2025analog,zhang2025toward}.
However, to the best of our knowledge, no prior work provides a dedicated security-focused treatment that is organized around the digital SemCom pipeline.

In practice, analog and digital SemCom differ in how semantic information is transmitted over the wireless interface, leading to distinct attack surfaces and design constraints. Analog SemCom typically maps semantic features directly to continuous channel inputs and relies on joint source-channel behavior. 
Digital SemCom, in contrast, transmits semantic information through discrete bits or symbols within practical digital transceiver pipelines.
In such designs, semantic information is mapped into a finite alphabet via explicit digital modulation, following two main routes: probabilistic modulation, which learns a conditional distribution and generates discrete symbols using neural network models, and deterministic modulation, which integrates explicit quantization.
Meanwhile, this discrete interface also facilitates the integration of cryptographic primitives.
This pipeline introduces additional vulnerabilities, including bit- or symbol-level semantic leakage and manipulation.
Moreover, digital SemCom also faces threats such as vulnerabilities within probabilistic or deterministic modulation components and attacks arising from packet-based delivery and protocol operations. 
These characteristics, together with the underexplored security risks in digital SemCom, motivate our dedicated focus on secure digital SemCom to provide a structured understanding of its threat landscape and corresponding defenses.

In this paper, we first introduce the fundamentals of SemCom, including the common architecture and the key differences between analog and digital SemCom. We then discuss the security and privacy threat landscape against digital SemCom, starting with challenges shared by both analog and digital SemCom and then focusing on those specific to digital implementations. Next, we analyze potential defense strategies against these threats. Finally, we outline open problems and future research directions toward secure and deployable digital SemCom systems.

\section{Background Knowledge of Semantic Communication}

\begin{figure*}[!t]
\begin{center}
\centerline{\includegraphics[width=1\linewidth]{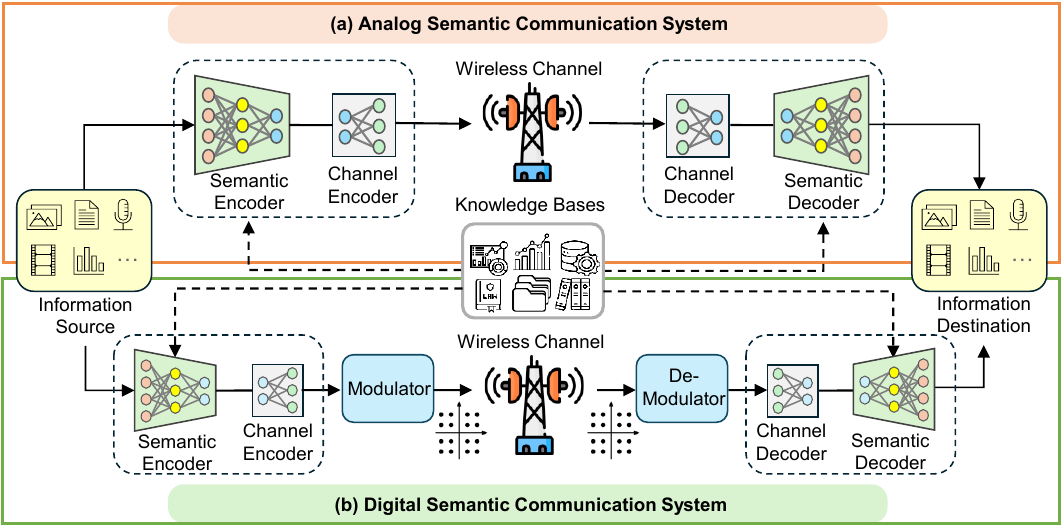}}
\caption{The end-to-end architectures of both analog and digital SemCom implementations: (a) analog SemCom framework; (b) digital SemCom framework. Architecturally, analog and digital SemCom mainly differ in whether the system explicitly defines a modulation-demodulation interface, as in standard digital transceivers, to transmit semantic information over a finite alphabet of discrete symbols or an equivalent bit stream.}
\label{fig.framework}
\end{center}
\vskip -0.1in
\end{figure*}

\begin{figure*}[!t]
\begin{center}
\includegraphics[width=0.95\linewidth]{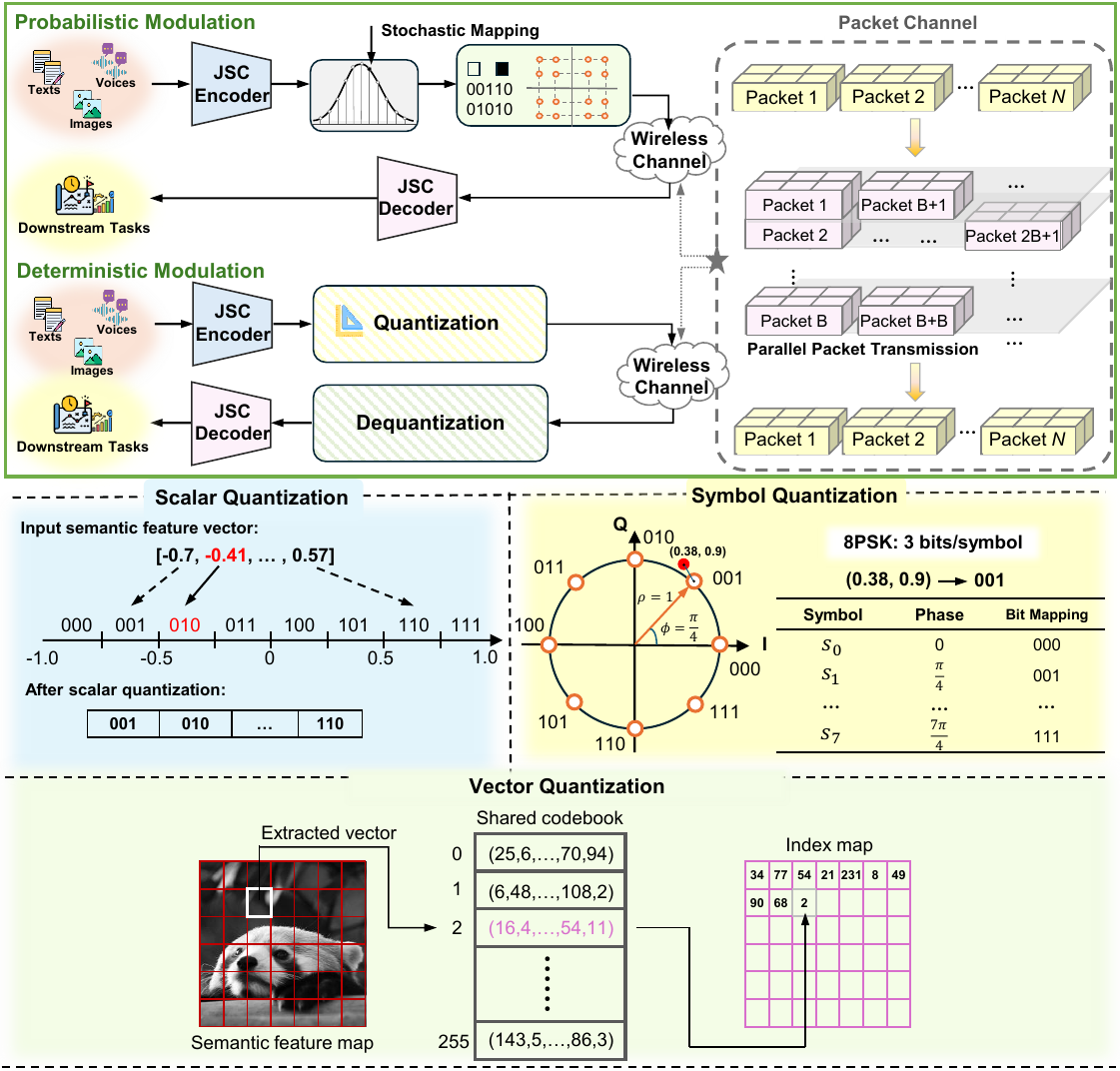}
\caption{
The detailed architecture of a digital SemCom system with explicit probabilistic or deterministic modulation and packet-based delivery. The JSC encoder/decoder denotes the joint source and channel coding/decoding module. Packet channel is a logical abstraction over the physical channel, capturing packetization and protocol effects.
}
\label{fig.digital_semcom_framework}
\end{center}
\vskip -0.1in
\end{figure*}

\begin{figure*}[!t]
\begin{center}
\includegraphics[width=0.9\linewidth]{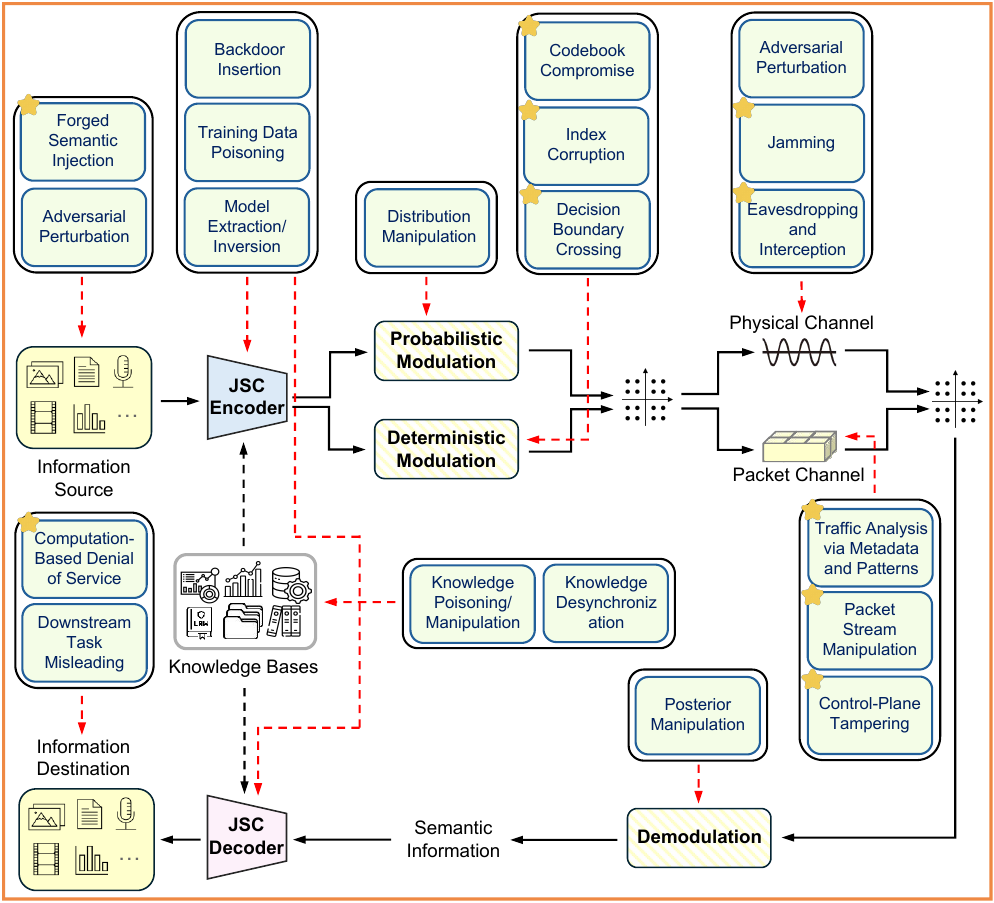}
\caption{
An overview of the security and privacy threats against digital SemCom. The red dashed lines link threats to the corresponding stages where they arise. For probabilistic modulation, demodulation is often integrated with the JSC decoder, whereas for deterministic modulation, demodulation and the JSC decoder are typically implemented as separate modules. For clarity and a unified presentation, we depict demodulation and the JSC decoder as two distinct blocks to highlight the threats targeting each stage. This overview includes threats shared by both analog and digital SemCom as well as those unique to digital SemCom. In particular, modulation-related threats and packet-channel threats are distinctive features of digital SemCom and are generally absent in analog implementations. In this paper, we primarily focus on threats specific to digital SemCom and their corresponding defenses.
Blocks with a star symbol in the upper-left corner denote threats that are also present in conventional digital communication systems but may be amplified in SemCom.
}
\label{fig.digital_semcom_threats}
\end{center}
\vskip -0.1in
\end{figure*}

The end-to-end architectures of analog and digital SemCom systems are shown in Fig.~\ref{fig.framework}.
The end-to-end SemCom framework typically includes an information source, semantic and channel coding modules, modulation and demodulation modules, the wireless channel, the corresponding semantic and channel decoding modules, an information destination, and a knowledge base that provides contextual support.
At the transmitter, the information source generates raw content such as text, speech, images, sensor measurements, or intermediate features produced by an edge model. A semantic encoder extracts and compresses task-relevant information from the source into semantic information, typically a low-dimensional feature or latent vector learned by neural networks. The semantic information is then processed by a channel encoder to improve robustness against channel impairments and support reliable delivery over the wireless channel.

The transmitter then modulates the semantic information into a transmitted signal, which propagates through the wireless channel and experiences fading, noise, and interference. 
At the receiver, the received signal is demodulated, and channel decoding mitigates channel distortions to recover reliable semantic information. The semantic decoder then reconstructs the intended meaning or directly outputs the task result for the information destination.
The information destination is defined at the receiver and reflects the target task. Throughout this process, a knowledge base can be maintained to store domain knowledge and contextual information, helping the transmitter and receiver align semantic interpretation and improve semantic consistency under limited communication resources.

Architecturally, analog and digital SemCom mainly differ in whether the system explicitly defines a digital modulation-demodulation interface, as in standard digital transceivers, to transmit semantic information over a finite alphabet. 
In analog SemCom, the learned semantic information is often mapped to the channel input in a continuous manner, and the receiver processes the received signal directly to recover semantic information or the task output. As a result, explicit analog modulation and demodulation modules are often absent or substantially simplified, since the end-to-end mapping already plays the role of converting semantic information into channel inputs and back.

In contrast, digital SemCom targets compatibility with practical physical-layer implementations, where transmission relies on a finite alphabet of discrete symbols or an equivalent bit stream. 
This discrete representation naturally supports packet-based delivery, where semantic units are segmented into packets and managed through protocol operations such as framing, sequencing, retransmission, and configuration exchange.
Accordingly, digital SemCom employs explicit digital modulation to map semantic information to a finite alphabet and carry it through practical digital transceiver pipelines.
At the receiver, the corresponding demodulation module maps the detected bits or symbols back to continuous semantic information, after which channel decoding and semantic decoding are applied to recover the intended meaning.
Specifically, Fig.~\ref{fig.digital_semcom_framework} illustrates the detailed architecture of a digital SemCom system.

Existing digital modulation designs for digital SemCom are commonly grouped into two paradigms \cite{zhang2025toward}: probabilistic modulation methods, which use probabilistic networks to generate discrete bits or symbols according to a learned distribution, and deterministic modulation methods, which apply direct quantization to map semantic information into discrete bits or symbols in a deterministic manner.

In probabilistic modulation \cite{bo2024deep}, the modulator is modeled as a stochastic mapping implemented by a neural encoder that outputs a conditional distribution over a finite bit or symbol alphabet. The transmitted sequence is generated by sampling from this learned distribution, while the demodulator is realized by a probabilistic inference network that estimates the posterior distribution of the transmitted bits or symbols given the received signal. A key challenge is that end-to-end training must handle discrete random variables, making posterior inference and gradient estimation nontrivial.
As a result, probabilistic digital SemCom often adopts techniques such as variational inference, score-function estimators, reparameterization-based approximations, and continuous relaxations such as Gumbel-Softmax. It also accounts for potential mismatch between relaxed training and discrete inference at deployment.

%

In deterministic modulation \cite{guo2024digital}, the system integrates an explicit quantizer-dequantizer pair that maps continuous semantic information into discrete representations and then reconstructs them for subsequent processing.
Depending on the design objective, common options include scalar quantization of individual feature elements followed by bit mapping and conventional digital modulation, symbol quantization that remaps learned symbol-like outputs onto a predefined constellation, and vector quantization with a shared codebook where semantic information is represented by codebook indices before bit mapping and conventional digital modulation.
Since quantization is not differentiable, deterministic schemes usually rely on differentiable approximations for end-to-end training, such as neural quantizer surrogates, straight-through estimators, soft-to-hard annealing, or additive-noise-based smoothing.

%

While conventional digital communication also involves modulation, digital SemCom changes the modulation interface because the transmitted bits or symbols are produced from learned semantic features and optimized for a downstream task. In probabilistic modulation, the transmitter outputs a conditional distribution over a discrete alphabet and generates the transmitted sequence by sampling, so an adversary can bias the recovered semantic information by manipulating the learned distribution or the sampling process. In deterministic modulation, discretization is realized via quantization, and in vector quantization variants the shared codebook and its indices become part of the semantic representation, so boundary crossing, index corruption, or codebook manipulation can cause systematic semantic remapping and task-level degradation.

\section{Security and Privacy Threats Against Digital Semantic Communication}

This section characterizes the security and privacy threat landscape against digital SemCom. We first summarize threats common to analog and digital SemCom, and then examine vulnerabilities specific to digital implementations, with emphasis on finite-alphabet semantic transmission and packet-based delivery.
An overview of the security and privacy threats against digital SemCom is illustrated in Fig.~\ref{fig.digital_semcom_threats}.
%
To clarify the boundary with conventional digital communication systems, in Fig.~\ref{fig.digital_semcom_threats}, blocks with a star symbol in the upper-left corner denote threats that are also present in conventional digital communication systems but may be amplified in SemCom. Blocks without the star symbol denote threats that are specific to SemCom systems.

\subsection{Common Challenges in Analog and Digital Semantic Communication}


Although this paper focuses on secure digital SemCom, it is helpful to begin with the security and privacy challenges shared by both analog and digital SemCom \cite{shen2023secure,yang2024secure}. 
These threats primarily arise from task-oriented semantic transmission and shared context, which can enable semantic leakage, semantic manipulation, and systematic bias through a compromised knowledge base or context. In addition, SemCom inherits model-related threats because semantic processing relies on machine learning, exposing the learning pipeline and trained models to attacks beyond the wireless channel. SemCom also faces authenticity and availability concerns.


Semantic leakage and inference attacks are among the most fundamental privacy challenges in SemCom.
Because the transmitted message is task-oriented semantic information, intercepted semantic information can already reveal sensitive attributes, intents, or environmental states.
The risk is amplified when the adversary combines the intercepted content with side information or public background knowledge.
SemCom is also vulnerable to semantic manipulation, where an attacker introduces carefully designed perturbations to distort the recovered meaning and mislead downstream tasks. Such attacks may not be reflected by conventional link-level metrics while still causing severe semantic errors. They also raise concerns about semantic integrity and source authenticity, since the receiver must determine whether the recovered meaning is reliable and originates from a legitimate transmitter rather than from fabricated or manipulated semantic information.

Another set of shared challenges arises from the supporting context and learning components. 
Many SemCom designs rely on shared context or knowledge base to align interpretation across endpoints.
This introduces risks such as knowledge poisoning, manipulation, and inconsistencies between transmitter and receiver knowledge states that can systematically bias semantic interpretation.
Moreover, since semantic extraction and reconstruction are typically implemented by machine learning models, SemCom is exposed to model-related threats throughout its lifecycle.
These threats include training data poisoning, backdoor insertion, and privacy attacks such as model extraction and inversion. 
Finally, SemCom remains exposed to availability threats, including jamming and flooding, as well as computation-based denial-of-service attacks targeting the semantic processing pipeline.
These attacks can degrade reliability and service continuity even when the physical link remains operational.

\subsection{Unique Challenges in Digital Semantic Communication}

\subsubsection{Attacks on Probabilistic Modulation}

Attacks on probabilistic modulation are more pronounced in digital SemCom because the transmitted representation is generated by sampling from a learned conditional distribution.
An adversary can target the distribution parameters to shift probability mass or alter distribution concentration, which biases the sampling behavior toward realizations that cause errors in the semantic decoding outcome, possibly in a targeted manner.

In the branch that targets discrete channel models such as the binary erasure channel (BEC) and binary symmetric channel (BSC), the encoder acts as a neural-network-parameterized Bernoulli generator whose outputs determine per-bit sampling probabilities. 
A primary threat is probability biasing, where an adversary induces structured perturbations to the source or intermediate semantic information, and thereby causes shifts in a subset of Bernoulli parameters.
As a result, the sampled bit patterns become more likely to trigger semantic decoding errors at the receiver.

In the constellation-symbol branch based on variational autoencoder (VAE)-style modeling, a key target is the probabilistic inference network at the receiver. 
By manipulating the inferred posterior so that probability mass concentrates on an incorrect hypothesis, or by inducing overconfident posteriors around the wrong mode, an adversary can cause systematic misinterpretation of semantic information.

\subsubsection{Attacks on Deterministic Modulation}
Deterministic modulation inserts a quantizer-dequantizer interface to convert continuous semantic information into discrete representations such as bits, symbols, or indices.
This discretization exposes hard decision boundaries and fixed mapping rules. As a result, small but structured perturbations near quantization or decision boundaries can be amplified into discrete changes that propagate through decoding and lead to erroneous semantic decoding outcomes.

In scalar quantization, an adversary can perturb selected feature elements to push them across quantization thresholds. Depending on the integer-to-bit mapping, such threshold crossings can induce multiple bit-level changes and thereby increase the likelihood of semantic decoding errors. 
In symbol quantization, remapping learned symbol-like outputs onto a predefined constellation introduces decision regions that can be exploited by carefully designed perturbations.
Such perturbations can shift received points across region boundaries and cause symbol decision errors.
In vector quantization with a shared codebook, semantic vectors are represented by codebook indices before bit mapping and conventional digital modulation. 
Perturbations near codeword boundaries can induce nearest-codeword confusion, and corruption of index bits can redirect decoding toward unintended codewords.
Because vector quantization relies on a shared codebook, codebook integrity and consistency between endpoints are critical.
Compromising the codebook through poisoning, unauthorized updates, or desynchronization can lead to systematic semantic misinterpretation.

\subsubsection{Packet- and Protocol-Level Threats}

Packet- and protocol-level threats are often more salient in digital SemCom because packet-level transport inherently operates on discrete representations.
In addition, digital SemCom may introduce semantic-related signaling and coordination state at the packet or control level.
This introduces privacy leakage beyond the semantic payload through observable traffic features such as packet sizes, timing, sequencing, and retransmission patterns. 
Additional leakage can arise from explicit semantic-related metadata carried in packets or control messages, including task identifiers, importance indicators, and version tags for models, codebooks, or knowledge bases. This metadata may enable an adversary to infer sensitive intent or context via traffic analysis.

Beyond passive leakage, adversaries can exploit packet and protocol mechanisms to induce semantic errors. General primitives such as injection, replay, reordering, fragmentation manipulation, and selective dropping can be amplified when semantic-aware packetization makes a small subset of high-impact packets dominate the semantic decoding outcome. Manipulating acknowledgments, retransmissions, and rate adaptation can further degrade semantic decoding quality, especially when feedback or control signaling is unauthenticated or weakly protected.


\section{Defenses against Attacks and Threats in Digital Semantic Communication}

\begin{figure*}[!t]
\begin{center}
\includegraphics[width=1\linewidth]{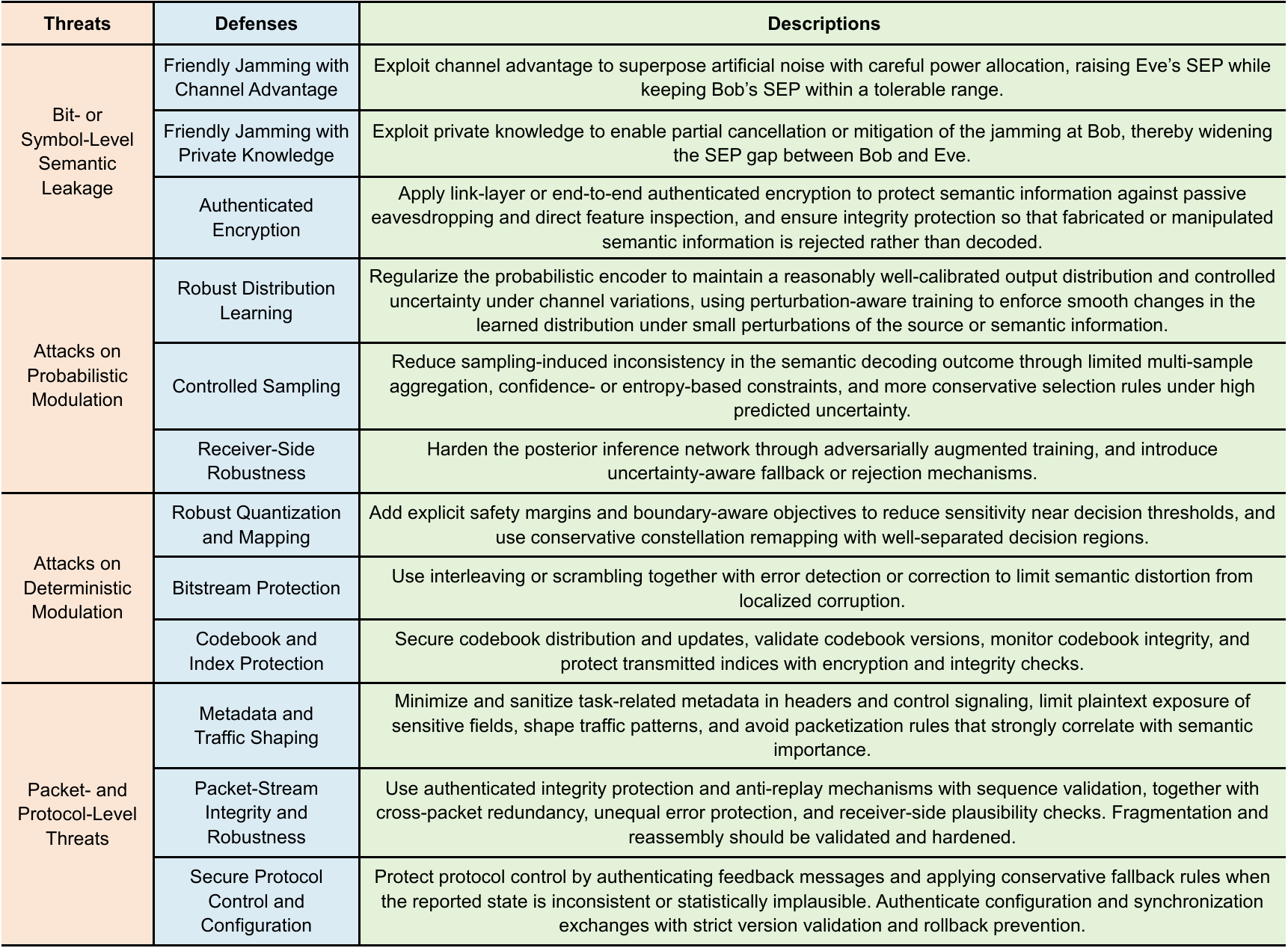}
\caption{
A summary table of possible defense strategies against security and privacy threats in digital SemCom, where Bob denotes the legitimate receiver and Eve denotes the eavesdropper.
}
\label{fig.digital_semcom_defenses_table}
\end{center}
\vskip -0.1in
\end{figure*}

This section analyzes defense strategies for mitigating security and privacy threats against digital SemCom. We structure the discussion by major attack surfaces, covering protections at the bit or symbol level, defenses for modulation-stage vulnerabilities, and countermeasures for packet- and protocol-level attacks in practical digital pipelines.
An overview of the corresponding defense strategies is illustrated in Fig.~\ref{fig.digital_semcom_defenses_table}.

\subsection{Defenses Against Bit- or Symbol-Level Semantic Leakage}

In digital SemCom, semantic information is transmitted through discrete symbols or an equivalent bit stream after modulation. This discrete interface provides a practical means for defending against semantic leakage at the physical layer. If an eavesdropper cannot reliably detect or decode the transmitted semantic symbols, the amount of usable task information that can be extracted is greatly reduced. Accordingly, an effective and practical defense objective is to keep the eavesdropper's symbol error probability (SEP) high, ideally close to the random-guessing regime \cite{chen2024nearly}. At the same time, the legitimate receiver should maintain a sufficiently low SEP so that semantic fidelity and downstream task performance remain acceptable.

SEPs can be derived or predicted from channel conditions, modulation order, coding rate, and transmit power, which makes SEP-based protection implementable at the physical layer. One representative mechanism is to superpose friendly jamming, also known as artificial noise (AN), onto the transmitted signal with careful power allocation \cite{chen2024nearly,chen2025knowledge}. The friendly jamming is designed to increase the eavesdropper's SEP while keeping the legitimate receiver's SEP within a tolerable range. This asymmetry can be achieved when the legitimate link has a channel advantage, such as stronger channel gain, better beamforming alignment, or more accurate channel knowledge, enabling the legitimate receiver to suppress or mitigate the interference more effectively than the eavesdropper.
The legitimate receiver may further benefit from private knowledge that is unavailable to the eavesdropper \cite{chen2025knowledge}. For example, it may have access to the friendly jamming source, the jamming sequence or coding rule, and the corresponding power allocation strategy, which enables partial cancellation or mitigation of the interference before decoding. This additional capability increases the SEP gap between the legitimate receiver and the eavesdropper, improving privacy protection while preserving the intended semantic task performance.

Importantly, the power allocation between the semantic information and friendly jamming can often be designed more systematically in digital SemCom than in analog settings. Because the relationship among coding, modulation, power allocation, channel conditions, and SEP is relatively well understood for practical digital transceivers, one can derive principled allocation schemes that satisfy reliability requirements for the legitimate receiver while degrading the eavesdropper. By contrast, in analog SemCom the link between continuous semantic distortions and information leakage is often harder to characterize, and friendly jamming strategies more often rely on numerical optimization or learning-based methods.

Besides physical-layer protection, encryption remains a fundamental and widely applicable tool for limiting semantic leakage in digital SemCom \cite{tung2023deep}. 
Since digital SemCom produces a discrete bit stream, conventional cryptographic primitives can be readily integrated.
For example, link-layer or end-to-end authenticated encryption can protect semantic information against passive eavesdropping and direct feature inspection.
Proper key establishment and management are essential.
Encryption is often most effective when combined with integrity protection, so that fabricated or manipulated semantic information is rejected rather than decoded.


\subsection{Defenses Against Attacks on Probabilistic Modulation}

Attacks on probabilistic modulation mainly aim to reshape the learned conditional distribution or to disrupt sampling behavior. 
A practical defense objective is therefore to make the transmitter distribution less sensitive to small perturbations in the source or intermediate semantic information. 
It should also prevent attack-induced inflation of predictive uncertainty or overconfident concentration that can amplify semantic errors.
%
This can be supported by regularizing the probabilistic encoder to maintain a reasonably well-calibrated output distribution and controlled uncertainty under channel variations.
In particular, we can adopt perturbation-aware training to enforce smooth changes in the learned distribution under small perturbations of the source or semantic information.
As a result, the distribution avoids large probability-mass shifts and abrupt concentration changes.
In addition, sampling-induced inconsistency in the semantic decoding outcome can be reduced by adopting controlled sampling at deployment.
For example, the receiver can draw a limited number of samples and aggregate them, enforce confidence- or entropy-based constraints, or switch to more conservative selection rules when predicted uncertainty is high.
These measures reduce the adversary's ability to induce unstable task outputs through distribution manipulation while introducing extra latency only when predicted risk is high.

For the bit-sequence branch with Bernoulli generators parameterized by neural networks, defenses should focus on limiting how much targeted perturbations can bias a subset of Bernoulli parameters and how strongly such biases propagate to erroneous decoder output and semantic decoding errors.
Practical measures include identifying Bernoulli probability dimensions that are unusually sensitive or semantically critical. Stronger protection or redundancy can then be applied to these dimensions to reduce the effect of probability biasing on sampled bit patterns.
For the constellation-symbol branch, the receiver typically relies on an explicit probabilistic inference network to form posterior beliefs over transmitted symbols. 
Robustness can be improved by hardening the posterior inference network against adversarial channel observations through adversarially augmented training and by introducing uncertainty-aware fallback or rejection mechanisms.
This helps reduce the likelihood that the receiver concentrates posterior mass on incorrect hypotheses or produces overconfident posteriors around a wrong mode under manipulated observations.
%

\subsection{Defenses Against Attacks on Deterministic Modulation}

Attacks on deterministic modulation are enabled by the explicit quantizer-dequantizer interface as well as the resulting hard decision boundaries and fixed mapping rules.
Defenses should therefore strengthen the discretization interface to reduce the likelihood that small feature perturbations change discrete outcomes, while improving the robustness of discrete representations as they traverse practical digital physical-layer procedures and packet-based delivery.
Practical measures include quantizer and mapping designs with explicit safety margins around decision thresholds, as well as boundary-aware objectives that discourage operating near decision thresholds.
They also include representation formats that limit the impact of localized errors, such as mappings with limited error propagation and lightweight error detection for discrete intermediates before semantic decoding.

%

For scalar quantization, defenses can emphasize robust bit mapping and bitstream protection.
Gray-style mappings reduce multi-bit changes caused by a single threshold crossing. Interleaving or scrambling, together with error detection or correction, can mitigate localized corruption before it causes large semantic distortions in downstream semantic decoding. 
For symbol quantization, robustness can be improved by conservative constellation remapping with sufficient separation between decision regions, and boundary-aware training to reduce sensitivity near decision thresholds.
For vector quantization, protection should secure codebook distribution and updates, validate codebook versions, and protect transmitted indices with encryption and integrity checks, since small index changes can redirect decoding to an unintended codeword and cause significant semantic deviation. Codebook integrity monitoring and authenticated updates further reduce the risk of poisoning or unauthorized modifications that manipulate codeword boundaries or endpoint consistency.
These defenses may constrain representation flexibility due to more conservative boundary designs. In addition, maintaining consistent codebook versions and authenticated updates across transmitter and receiver increases system-level coordination requirements.

\begin{figure*}[!t]
\begin{center}
\includegraphics[width=0.93\linewidth]{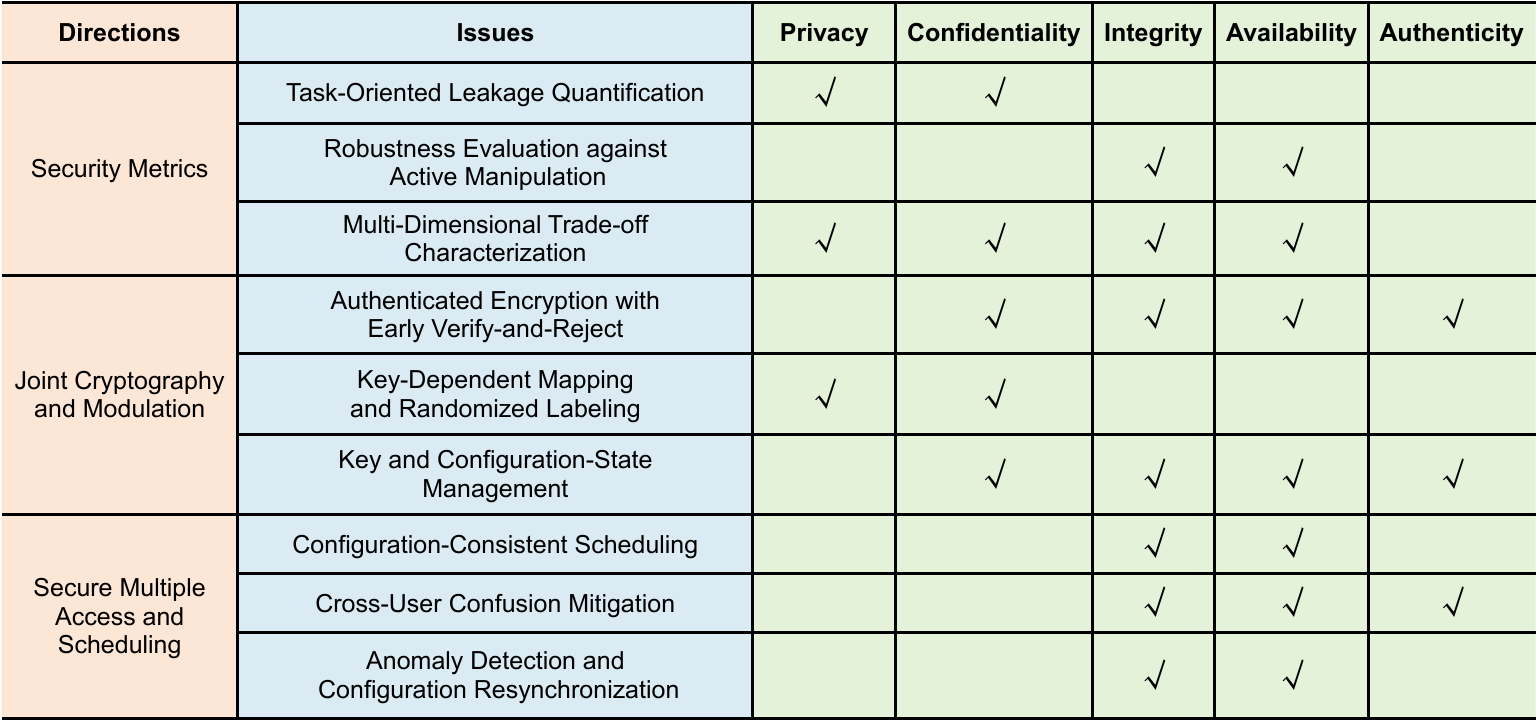}
\caption{An overview of open problems and future research directions for secure and deployable digital SemCom.}
\label{fig.digital_semcom_future}
\end{center}
\vskip -0.1in
\end{figure*}


\subsection{Defenses Against Packet- and Protocol-Level Threats}

Packet- and protocol-level defenses in digital SemCom should pursue two complementary goals: reducing side-channel leakage introduced by packetization and metadata, and preventing active manipulation of packet delivery and protocol state.
To improve privacy, task-related metadata in headers and control signaling should be minimized and sanitized.
In addition, sensitive fields should not be exposed in plaintext whenever possible, such as task identifiers, importance labels, model or codebook versions, and knowledge-base synchronization markers.
%
%
Resistance to traffic analysis can also be improved by shaping observable patterns through padding, batching, or controlled scheduling.
In addition, packetization rules that strongly correlate with semantic importance should be avoided. Otherwise, these correlations may leak packet importance via packet sizes, timing, or retransmission patterns.

To mitigate active attacks, digital SemCom can rely on standard packet security and protocol hardening, while accounting for the semantic setting where a small subset of packets may dominate task performance.
Authenticated integrity protection defends against injection and manipulation, and anti-replay mechanisms with sequence validation mitigate replay and reordering for both payload packets and semantic-related control messages. 
Robustness to selective dropping can be improved through cross-packet redundancy and unequal error protection for task-critical packets. In addition, receiver-side plausibility checks can detect inconsistencies in recovered semantic information under missing or altered packets, particularly when high-impact packets are selectively removed. 
Fragmentation and reassembly should be validated and hardened. This prevents malformed fragments and ambiguous reassembly outcomes from corrupting semantic information or bypassing sequence validation and integrity checks.
Protocol operations such as acknowledgments, retransmissions, and rate adaptation should be protected against forged feedback. This can be achieved by authenticating feedback messages and applying conservative fallback rules when the reported state is inconsistent or statistically implausible.
Configuration and synchronization exchanges should be authenticated, including coding and modulation settings, semantic model versions, codebooks, and knowledge base updates.
Strict version validation and rollback prevention are also needed. 
These measures help avoid persistent transmitter-receiver mismatch that can degrade semantic decoding quality over multiple packets.
While these mechanisms improve robustness, padding and redundancy may increase transmission delay. Protecting acknowledgments, retransmissions, and configuration exchanges also adds complexity to protocol control and state management.

\section{Open Problems and Future Research Directions}

This section discusses several open problems and future research directions that we consider important for secure and deployable digital SemCom. We focus on open issues in evaluation methodology and system design, and outline promising paths toward practical security mechanisms and real-world deployment.
An overview of the open problems and future research directions is illustrated in Fig.~\ref{fig.digital_semcom_future}.

\subsection{Security Metrics for Digital SemCom}

A key future direction is to establish security metrics tailored to digital SemCom, since conventional link metrics do not directly reflect semantic leakage, semantic integrity, or privacy risk. Evaluation should quantify an adversary's capability in task-relevant terms, for example what sensitive attributes or intents can be inferred from intercepted semantic traffic and what task-level advantage an attacker can obtain. It should also measure robustness to active manipulation by assessing how targeted bit-, symbol-, or packet-level attacks affect recovered meaning and downstream task performance. These security metrics should be reported jointly with semantic utility and communication cost to reveal key tradeoffs among privacy, integrity, reliability, and efficiency.

\subsection{Joint Cryptography and Modulation for Secure Digital SemCom}

Since digital SemCom transmits semantic information through bits or symbols, security can be co-designed with modulation and demodulation, rather than applied as an external layer. In particular, encryption and authentication can be coupled with the semantic-to-symbol mapping, framing, and receiver detection so that confidentiality and authenticity are enforced at the same discrete interface where semantic information is carried. Representative directions include authenticated encryption that supports early verification and rejection of injected or replayed semantic packets before semantic decoding, and key-dependent symbol mappings or randomized labeling to reduce repeatable patterns that enable probing and inference. Practical co-design should also consider key establishment and synchronization, protected signaling for semantic model or codebook state, and the tradeoff between security overhead and link reliability.

\subsection{Secure Multiple Access and Scheduling for Discrete Semantic Streams}

In digital SemCom, meaning is carried as packet-based bits or symbols, and correct interpretation often depends on shared semantic-side configuration, such as the semantic model version, codebook, or task context agreed between the transmitter and receiver.
This makes multiuser operation more tightly coupled to semantic information than in conventional data networks.
Collisions, imperfect multiuser separation, or mis-scheduling can cause bit or symbol sequences to be processed under the wrong semantic context or configuration.
This can lead to semantic ambiguity even when user identity is correctly established at the packet level and the physical link remains usable.
Adversaries can exploit this coupling in several ways, such as manipulating contention and scheduling to delay or desynchronize configuration updates, injecting packets that imitate another user's semantic format, or selectively disrupting packets that carry configuration negotiation.
As a result, subsequent packets may be processed under mismatched models or codebooks.
These risks motivate the design of multiple access and scheduling mechanisms for discrete semantic pipelines.
Such mechanisms may include scheduling policies that maintain consistent configuration across users and time, safeguards that reduce cross-user confusion at the bit or symbol stream level, and recovery procedures that restore consistency when anomalies or desynchronization are detected.

\section{Conclusion}

In this paper, we investigated security and privacy issues in digital SemCom. We reviewed fundamental SemCom concepts and architectures and clarified key differences between analog and digital SemCom. We analyzed major threats against digital SemCom, covering semantic-layer vulnerabilities as well as those introduced by finite-alphabet semantic transmission, modulation, and packet-based delivery procedures. Based on this threat analysis, we discussed defense strategies for mitigating these security and privacy threats and outlined future research directions. Digital SemCom is a promising enabler for future intelligent wireless networks, and effective security and privacy protections are essential to make it trustworthy and deployable in real-world systems.

\bibliographystyle{IEEEtran}

\bibliography{IEEEabrv,myref}

\section*{Biographies}

\begin{IEEEbiographynophoto}{Weixuan Chen}
(Graduate Student Member, IEEE) received the B.E. degree from Sichuan University, Chengdu, China, in 2022. 
He is currently working toward the Ph.D. degree with the College of Information Science and Electronic Engineering, Zhejiang University, Hangzhou, China. 
His research interests include semantic communications and computer vision. 
He was selected for the China Association for Science and Technology (CAST) Young Scientific and Technological Talent Support Program (Doctoral Student Special Program) in 2025. 
He received the Best Paper Award at the 2025 IEEE International Conference on Communications in China (ICCC).
\end{IEEEbiographynophoto}

\begin{IEEEbiographynophoto}{Qianqian Yang}
(Member, IEEE) received the B.Sc. degree in automation from Chongqing University, Chongqing, China, in 2011, the M.S. degree in control engineering from Zhejiang University, Hangzhou, China, in 2014, and the Ph.D. degree in electrical and electronic engineering from the Imperial College London, U.K. In 2016, she held visiting positions with CentraleSupelec. She was also with New York University Tandon School of Engineering from 2017 to 2018. She was a Post-Doctoral Research Associate with the Imperial College London and a Machine Learning Researcher with Sensyne Health Plc. She is currently a tenure-tracked Professor with the Department of Information Science and Electronic Engineering, Zhejiang University. Her research interests include wireless communications, information theory, and semantic communications.
\end{IEEEbiographynophoto}

\begin{IEEEbiographynophoto}{Yuanyuan Jia}
(Graduate Student Member, IEEE) received the B.E. degree from Sun Yat-sen University, Guangzhou, China, in 2025. 
She is currently working toward the M.E. degree with the Polytechnic Institute of Zhejiang University, Hangzhou, China.
\end{IEEEbiographynophoto}

\begin{IEEEbiographynophoto}{Junyu Pan} (Graduate Student Member, IEEE) received the B.E. degree from Zhejiang University, Hangzhou, China, in 2025. 
He is currently working toward the M.E. degree with the College of Information Science and Electronic Engineering, Zhejiang University, Hangzhou, China.
\end{IEEEbiographynophoto}

\begin{IEEEbiographynophoto}{Shuo Shao} (Member, IEEE) received the B.S. degree in information science from Southeast University, Nanjing, China, in 2011, the M.A.Sc. degree in electrical and computer engineering from McMaster University, Hamilton, ON, Canada, in 2013, and the Ph.D. degree from Texas A\&M University, College Station, TX, USA, in 2017. He is currently an Associate Professor with the Department of System Science, University of Shanghai for Science and Technology, and was an Associate Professor with the School of Electronics Information and Electrical Engineering, Shanghai Jiao Tong University, China. His research interests include network information theory, algebraic code, machine learning, and semantic communications.
\end{IEEEbiographynophoto}

\begin{IEEEbiographynophoto}{Jincheng Dai}(Member, IEEE) received the B.S. and Ph.D. degrees from the Beijing University of Posts and Telecommunications (BUPT), Beijing, China. He is currently an Associate Professor with the School of Artificial Intelligence and the Key Laboratory of Universal Wireless Communications, Ministry of Education, BUPT. His research interests include AI for coding and communications. He has been selected for the Beijing Nova Program, the China Association for Science and Technology's (CAST) Young Elite Scientists Sponsorship Program, and the Xiaomi Young Scholars Program. As a core contributor, he has received the First-Class Natural Science Award of the Chinese Institute of Electronics. He has received several accolades, such as the Excellent Science and Technology Paper Award from CAST.
\end{IEEEbiographynophoto}

\begin{IEEEbiographynophoto}{Meixia Tao} (Fellow, IEEE) received the B.S. degree in electronic engineering from Fudan University, Shanghai, China, in 1999, and the Ph.D. degree in electrical and electronic engineering from The Hong Kong University of Science and Technology in 2003. She is currently a Professor with the School of Information Science and Electronic Engineering, Shanghai Jiao Tong University, China. Her current research interests include wireless edge learning, semantic communications, integrated communication-computing-sensing, and AI-based channel modeling and beamforming. She was a recipient of the 2019 IEEE Marconi Prize Paper Award, the 2013 IEEE Heinrich Hertz Award for Best Communications Letters, the IEEE/CIC International Conference on Communications in China (ICCC) 2015 Best Paper Award, and the International Conference on Wireless Communications and Signal Processing (WCSP) 2022 and 2012 Best Paper Awards. She also received the 2009 IEEE ComSoc Asia-Pacific Outstanding Young Researcher award.
\end{IEEEbiographynophoto}

\begin{IEEEbiographynophoto}{Ping Zhang}
(Fellow, IEEE) is currently a professor of School of Information and Communication Engineering at Beijing University of Posts and Telecommunications, the director of State Key Laboratory of Networking and Switching Technology, a member of IMT-2020 (5G) Experts Panel, a member of Experts Panel for China’s 6G development. He served as Chief Scientist of National Basic Research Program (973 Program), an expert in Information Technology Division of National High-tech R\&D program (863 Program), and a member of Consultant Committee on International Cooperation of National Natural Science Foundation of China. His research interests mainly focus on wireless communication. He is an Academician of the Chinese Academy of Engineering (CAE).
\end{IEEEbiographynophoto}

\vspace{12pt}

\end{document}